\documentclass[conference]{IEEEtran}
\IEEEoverridecommandlockouts
\usepackage[compress,sort]{cite}
\usepackage{amsmath,amssymb,amsfonts}
\usepackage{algorithmic}
\usepackage{graphicx}
\usepackage{multirow}
\usepackage{comment}
\usepackage{balance}
\usepackage{textcomp}
\usepackage{float}
\usepackage{xcolor}
\usepackage{amsmath}
\def\BibTeX{{\rm B\kern-.05em{\sc i\kern-.025em b}\kern-.08em
    T\kern-.1667em\lower.7ex\hbox{E}\kern-.125emX}}
\begin{document}

\title{Transfer Learning-Based Deep Residual Learning for Speech Recognition in Clean and Noisy Environments}

\makeatletter 
\newcommand{\linebreakand}{%
  \end{@IEEEauthorhalign}
  \hfill\mbox{}\par
  \mbox{}\hfill\begin{@IEEEauthorhalign}
}
\makeatother 

\author{\IEEEauthorblockN{ Noussaiba Djeffal}
\IEEEauthorblockA{\textit{Speech and Signal Processing Laboratory} \\
\textit{University of Sciences and Technology, USTHB}\\
Algiers, Algeria\\
ndjeffal@usthb.dz}
\and
\IEEEauthorblockN{ Djamel Addou}
\IEEEauthorblockA{\textit{Speech and Signal Processing Laboratory} \\
\textit{University of Sciences and Technology, USTHB}\\
Algiers, Algeria\\
daddou@usthb.dz }
\and 
\linebreakand
\IEEEauthorblockN{ Hamza Kheddar}
\IEEEauthorblockA{\textit{LSEA Laboratory, dept. Electrical engineering} \\
\textit{ University of MEDEA}\\
Medea, Algeria \\
kheddar.hamza@univ-medea.dz}
\and
\IEEEauthorblockN{ Sid Ahmed Selouani}
\IEEEauthorblockA{\textit{Research Laboratory in Human-System Interaction} \\
\textit{Universite de Moncton, Shippagan Campus}\\
Shippagan, Canada\\
 sid-ahmed.selouani@umoncton.ca}
}

\makeatletter

\def\ps@headings{%
\def\@oddhead{\parbox[t][\height][t]{\textwidth}{\flushleft

\noindent\makebox[\linewidth]
}
\vspace{0.5cm}
\hfil\hbox{}}%
\def\@oddfoot{\MYfooter}%
\def\@evenfoot{\MYfooter}}

\def\ps@IEEEtitlepagestyle{%
\def\@oddhead{\parbox[t][\height][t]{\textwidth}{
2024 International Conference on Telecommunications and Intelligent Systems (ICTIS)\\

}\hfil\hbox{}}%

\def\@oddfoot{ 979-8-3315-2739-6/24/\$31.00 \textcopyright 2024 IEEE \hfil 
\leftmark\mbox{}}%
\def\@evenfoot{\MYfooter}}

\maketitle

\begin{abstract}
Addressing the detrimental impact of non-stationary environmental noise on automatic speech recognition (ASR) has been a persistent and significant research focus. Despite advancements, this challenge continues to be a major concern. Recently, data-driven supervised approaches, such as deep neural networks, have emerged as promising alternatives to traditional unsupervised methods. With extensive training, these approaches have the potential to overcome the challenges posed by diverse real-life acoustic environments. In this light, this paper introduces a novel neural framework that incorporates a robust front-end into ASR systems in both clean and noisy environments. Utilizing the Aurora-2 speech database, the authors evaluate the effectiveness of an acoustic feature set for Mel-frequency, employing the approach of transfer learning based on Residual neural network (ResNet). The experimental results demonstrate a significant improvement in recognition accuracy compared to convolutional neural networks (CNN) and long short-term memory (LSTM) networks. They achieved accuracies of 98.94\% in clean and 91.21\% in noisy mode.  \\
\end{abstract}


\section{Introduction}
In recent years, automatic speech recognition (ASR) has made significant strides, leading to notable advancements in performance. These advancements have facilitated the integration of speech-specific intelligent human-machine communication systems, like smartphone assistants (e.g., Cortana, Siri) \cite{li2022recent}, and biomedical application \cite{essaid2024enhancing,essaid2025deep}. However, despite these achievements, a fundamental challenge remains: the degradation of their performance in everyday situations caused by ambient noise and reverberation, which adversely affect the captured speech signals received by the microphones. The last few years, have witnessed significant advancements in deep neural networks, which have played a central role in recent developments  \cite{kheddar2024automatic}. It has proven to be a powerful approach for leveraging vast amounts of training data to build complex and specialized analysis systems \cite{zhang2017advanced}, and has achieved significant success in diverse fields such as understanding and generating human language  \cite{djeffal2023automatic}, biomedical such as severe hearing loss \cite{essaid2024advanced}, speech security  \cite{noureddine2023adversarial}, among others. These achievements have spurred increased research efforts in deep learning to enhance the robustness of ASR in noisy environments \cite{djeffal2023noise}.

Since the introduction of ResNet \cite{he2016deep}, the incorporation of residual connections has become a cornerstone in many leading neural network architectures. This innovation has spurred a series of breakthroughs across various domains, including computer vision and data mining. Empirical evidence consistently shows that residual connections greatly alleviate the challenge of training deep neural networks to fit the training data while preserving excellent generalization capabilities on test datasets \cite{he2020resnet}. Despite these empirical successes, there has been limited theoretical analysis regarding the impact of residual connections on the generalization ability of deep neural networks for ASR in noisy environments. Addressing the detrimental impact of non-stationary environmental noise on ASR has been a persistent and significant research focus. Despite advancements, this challenge continues to be a major concern. Recently, to address these problematic, a wave of research efforts has emerged to address the challenges associated with robust speech recognition, as evidenced by the REVERB and CHiME challenges  \cite{barker2013pascal,jalalvand2015boosted,kinoshita2016summary}. Inspired by these endeavors, our work aims to leverage large-scale training data to achieve cleaner signals and features from noisy speech audio or directly perform recognition of noisy speech, specifically in the multicondition setting of Aurora-2, such as  \cite{nasret2021design} in this paper, the authors design enhancements and modifications for automatic speech recognition systems in noisy environments, in   \cite{soe2020discrete} the paper integrates discrete wavelet denoising into MFCC for noise suppression in automatic speech recognition systems using the Aurora-2 dataset. This paper's contributions are categorized into three key areas:
\begin{itemize}
   
    \item Propose an extension architecture (target model) for pretrained ResNet model (source model), to learn both clean and noisy modes.
     \item  Use two modes: clean speech and noisy speech, with four noise scenarios (suburban train, babble, car, and exhibition hall) at different signal-to-noise ratios (SNRs).
    \item Compare the results of ResNet with CNN, LSTM, BiLSTM, and a concatenated CNN-LSTM, as employed in \cite{gueriani2024enhancing}, in both clean and noisy modes.
\end{itemize}
The structure of this article is outlined as follows: Section II covers the background of transfer learning-based deep residual learning. Section III provides a brief summary of related works. Section IV presents a comprehensive summary of the proposed ResNet  model for speech recognition systems. The experimental investigations are detailed in Section V, which includes two case studies conducted with the proposed model. Finally, Section VI concludes the article.

\section{Background}
Transfer learning is a commonly used approach in deep learning that involves adapting a model pre-trained on a broad and general dataset, such as ImageNet, to a more specific task. In the realm of ASR, models like ResNet, VGG16 \cite{tsalera2021comparison}, and others, originally trained on ImageNet, can be repurposed to handle speech data by using spectrograms, which are visual representations of audio signals  \cite{rebouh2023blind,lachenani2024improving}, as inputs. These models, initially designed to capture a wide array of visual features from images, are particularly advantageous for ASR tasks with limited labeled data. Typically, the early layers of the pre-trained model, which learn to recognize basic visual patterns like edges and textures, are retained, while the later, more specialized layers are fine-tuned or replaced to better fit the ASR task. For instance, a VGG16 model pre-trained on ImageNet can be adapted and fine-tuned to recognize speech in noisy environments, utilizing the previously learned features to enhance accuracy and reduce the training time required for speech-related tasks.

In practice, transfer learning generally involves several key steps. Initially, a model is trained on a large dataset from a source domain, which typically involves a related task. The next step is to fine-tune this pre-trained model on a smaller dataset from the target domain. Fine-tuning can range from retraining the entire model to adjusting only a few layers, based on the similarity between the source and target tasks. There are several techniques for adapting models through transfer learning. Feature extraction uses the pre-trained model to derive features from the new data, which are then fed into a separate model specifically trained for the target task. In this approach, most of the pre-trained model’s layers are kept unchanged, with only the final layers being retrained. Fine-tuning involves adjusting the pre-trained model by retraining some or all of its layers with a smaller learning rate, which is effective when the target task closely aligns with the source task. This allows the model to adapt to new data while retaining the valuable features learned previously.  This technique preserves useful general features while tailoring specific layers to better fit the new task \cite{zhuang2020comprehensive}.

The pre-trained model ResNets  Introduced by He et al. in 2015 \cite{sankupellay2018bird} have been a significant advancement in the field of deep learning, particularly for training very deep networks. It address the issue of vanishing gradients, which often hampers the training of deep neural \cite{zhang2023transfer}.

The key innovation in ResNet is the introduction of residual blocks, in contrast to a traditional neural network \cite{kheddar2023deep} where each layer feeds into the next. However, in a residual block, the input to a layer is also fed directly into a layer deeper in the network. This skip connection or shortcut allows the model to learn residual functions with reference to the layer inputs, rather than learning unreferenced functions directly. Essentially, instead of trying to learn the output directly, the network learns the difference (residual) between the input and the output, which simplifies the learning process and helps mitigate the vanishing gradient problem. There are different versions of ResNet, such as ResNet-18, ResNet-34, ResNet-50 \cite{mazari2023deepTL}, ResNet-101, and ResNet-152, where the numbers denote the number of layers. ResNet-50, for example, uses 50 layers and incorporates both identity and convolutional blocks. The identity blocks keep the same input and output dimensions, while the convolutional blocks change the dimensions using convolutional layers.

\section{Related research}
ResNet had received significant attention from researchers in noisy speech such as   \cite{yang2023resnet} the authors in this study uses a three-feature fusion method called Net50-SE, combining a deep residual network (ResNet) with an attention mechanism. The ResNet structure extracts features from environmental sound signals, while the attention module focuses on important feature map channels and suppresses environmental noise, improving classification accuracy, in \cite{mohammadamini2022learning} paper proposes two new ResNet-based speaker recognition systems that enhance robustness against additive noise and reverberation. These systems aim to extract x-vectors in noisy environments that closely match those in clean environments by jointly minimizing speaker classification loss and the distance between noisy and clean x-vectors. The modified systems are tested under various noise and reverberation conditions, demonstrating improved efficiency. In clean speech such as \cite{alzantot2019deep} this paper addresses threats to automatic speaker verification systems from synthetic speech and replay attacks by developing three variants of residual convolutional networks for the ASVSpoof2019 competition, the authors in  \cite{saleem2022e2e} use deep residual convolutional neural networks for end-to-end video  driven speech synthesis. This study   \cite{ashurov2022environmental} explores the effectiveness of pre-trained CNNs for environmental sound classification by converting raw audio signals into log-Mel spectrograms. The paper evaluates various hyperparameters and optimizers, such as Adam and RMSprop, on models including Inception, VGG, ResNet, and DenseNet201. Using the UrbanSound8K dataset, DenseNet201 achieved 97.25\% accuracy, while ResNet50V2 achieved 95.5\%, demonstrating high performance in audio categorization. In \cite{tang2018acoustic} the paper presents DenseRNet, a novel model for multichannel speech recognition that combines DenseNet and ResNet features. DenseRNet improves gradient flow and multi-resolution feature utilization surpassing the baseline method.

\section{Proposed approach}

ResNets employ skip connections to bypass one or more layers, typically including non-linearities similar to the use of the rectifier activation function (ReLU) and batch normalization, this technique is applied between skipped layers. It effectively tackles the issues of vanishing gradients and accuracy saturation that can occur when adding more layers to a deep model, which can lead to increased training error \cite{zhang2017advanced}.
Consider \( z \) as the input. As \( z \) passes through various layers in each block of the ResNet cell, such as convolutional, dropout, and normalization layers, the output \( y \) is produced. 
The loss function is evaluated using the input \( z \), with \( y \) defined as \( f(z) \), where \( f(z) \) represents the loss function. By incorporating a skip connection, the input \( z \) is combined with the output, resulting in \( y = f(z) + z \). The aim is for \( f(z) \) to diminish towards zero, which enables the network to learn effectively from the discrepancy between the input and the output \cite{reza2023customized}.

\begin{figure}[htbp]
\centering
\includegraphics[width=0.5\textwidth,height=0.6\textheight,keepaspectratio]{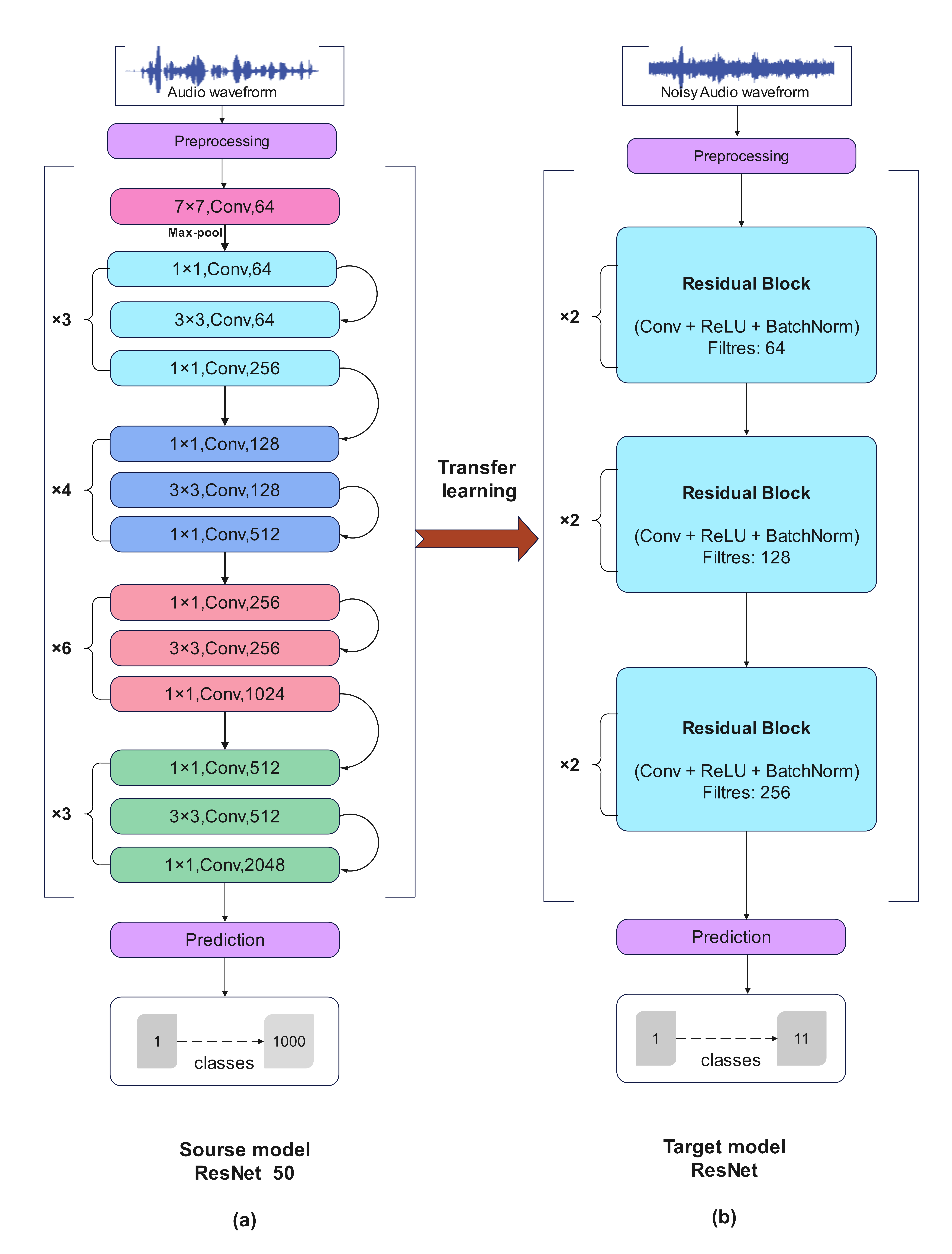}
    \caption{Proposed scheme:  (a) Source model \cite{zhang2023transfer} (b) Target model.}
   \label{figure1}
\end{figure}

\subsection {Description of the source model (before Transfer learning)}

ResNet-50 is a deep convolutional neural network architecture widely used for speech recognition tasks. It begins with a preprocessing step to prepare the input, followed by a 7x7 convolutional layer with 64 filters and max-pooling. The core of ResNet-50 consists of four stages of residual blocks, each designed to extract increasingly complex features. The first stage has three residual blocks, each with a 1x1 convolution with 64 filters, a 3x3 convolution with 64 filters, and a 1x1 convolution with 256 filters. The second stage contains four residual blocks with 1x1 convolutions of 128 filters, 3x3 convolutions of 128 filters, and 1x1 convolutions of 512 filters. The third stage includes six residual blocks, each with 1x1 convolutions of 256 filters, 3x3 convolutions of 256 filters, and 1x1 convolutions of 1024 filters. The fourth stage has three residual blocks with 1x1 convolutions of 512 filters, 3x3 convolutions of 512 filters, and 1x1 convolutions of 2048 filters. Each block incorporates skip connections that help mitigate the vanishing gradient problem, enabling deeper networks to be trained effectively. Finally, a prediction layer produces a probability distribution over 1000 classes. This architecture is designed to ease the training of deep networks and improve their performance on various computer vision tasks. The ResNets-50 architecture is in figure \ref{figure1} (a).

\subsection {Description of the target model (after Transfer learning)}

The proposed ResNet model for classification speech is outlined as follows: the model begins with an input layer designed to match the shape of the training data. The input passes through an initial convolutional layer with 64 filters, a kernel size of 3, and a ReLU activation function, followed by a max-pooling layer with a pool size of 2. Next, the input undergoes three residual blocks, each consisting of two convolutional layers (with batch normalization and ReLU activation) and a shortcut connection that matches the number of filters through a convolution. After each residual block, another max-pooling layer reduces the spatial dimensions. Specifically, the first residual block has 64 filters, the second has 128 filters, and the third has 256 filters. The output from the final residual block is flattened and passed through a dense layer with 128 units and ReLU activation. A dropout layer with a 0.5 dropout rate follows to prevent overfitting. The model concludes with an output dense layer using a softmax activation function to classify  the input  11 classes into one of several classes. The ResNets architecture is illustrated in Figure \ref{figure1} (b).


\section{Experiments}

\subsection{Aurora dataset}\label{AA}
The Aurora 2 database, developed for the ETSI STQ-AURORA DSR working group, supports the assessment of speech recognition algorithms in noisy conditions. It is based on the TIDigits dataset, featuring eleven connected digit utterances from American English speakers, downsampled to 8kHz 

The database offers two training modes: one with clean data and another with a mix of clean and noisy data. The clean mode includes 8,440 utterances filtered with G.712 characteristics, while the multi-condition mode divides the same utterances into 20 subsets, representing four noise scenarios (suburban train, babble, car, exhibition hall) at five SNR levels (20dB, 15dB, 10dB, 5dB, clean).

Three test sets were developed using 4,004 utterances, equally sourced from 52 women and 52 men, each containing 1,001 utterances. Noise was introduced at SNRs ranging from 20dB to -5dB, with additional clean conditions. Test sets include various environmental noises, such as subway stations, cars, restaurants, and train stations \cite{djeffal2023noise}.

\subsection{ Experimental results of ResNet}

Table \ref{tab01} shows the accuracy of ResNet before and after transfer learning in both clean and noisy modes. It demonstrates high accuracy of ResNet after transfer learning. 

\renewcommand{\arraystretch}{1.5} 
 \begin{table}[h]
\centering
\caption{Accuracy for ResNet before and after Transfer learning.}
\label{tab01}
\begin{tabular}{|l|l|l|}
\hline
\textbf{ResNet }             & \textbf{Clean Speech}   & \textbf{Noisy Speech}  \\ \hline
ResNet before Transfer learning                  & 94.54\% & 83.43\%\\ 
\textbf{ ResNet after Transfer learning}              & \textbf{98.94}\% &\textbf{ 91.21}\% \\ \hline
\end{tabular}
\end{table}

\subsection{Analysis of results }

 The Aurora2 dataset was utilized, containing a total of 4,824 isolated digit files, with an equal distribution of 2,412 files for clean mode and 2,412 for noisy mode. Approximately 40\% of the data was used for testing. The multi-condition mode included four noise scenarios (suburban train, babble, car, and exhibition hall) at five different SNRs: 20dB, 15dB, 10dB, 5dB, and clean condition. The dataset encompasses 11 classes, ranging from 0 to 'oh'. Figure \ref{figure2} of confusion matrix for the 11-class multiclass classification further validates these findings, and Figure  \ref{figure3} illustrate a confusion matrix of binary classification clean and noisy.

\renewcommand{\arraystretch}{1.15} 
 \begin{table}[h]
\centering
\caption{Accuracy for models tested in clean and  multi-condition training.}
\label{tab02}

\begin{tabular}{|l|l|l|}
\hline
\textbf{Models Tested }             & \textbf{Clean Speech}   & \textbf{Noisy Speech}  \\ \hline
CNN                  & 97.21\% & 90.12\%\\ 
LSTM                 & 96.06\% & 86.12\% \\
BiLSTM               & 94.33\% & 83.43\% \\
Concatenate LSTM-CNN & 97.96\% & 90.72\% \\\textbf{ResNet }              & \textbf{98.94}\% &\textbf{ 91.21}\% \\ \hline
\end{tabular}
\end{table}

Table \ref{tab02} presents the accuracy of the models tested in both clean and multi-condition modes. It shows the recognition rates obtained from experiments that used stochastic gradient descent (SGD) optimizers with a learning rate of 0.001. The experiment demonstrates that the system designed with ResNet achieves significantly higher recognition rates compared to CNN, LSTM, BiLSTM, and CNN-LSTM. As shown in Table \ref{tab01}, in clean mode ResNet achieves a recognition rate of 98.94\%, surpassing the rates of CNN, LSTM, BiLSTM, and CNN-LSTM. In a noisy environment, ResNet also performs better, with a recognition rate of 91.21\%, which is higher than CNN, LSTM, BiLSTM, and CNN-LSTM.

In Figure \ref{figure4}, the word error rate (WER) in a clean environment indicate that ResNet performs moderately better than the other methods. However, in a noisy environment, ResNet achieves significantly lower WER compared to CNN, LSTM, BiLSTM, and CNN-LSTM.

\begin{figure}[htbp]
\centering
\includegraphics[width=0.5\textwidth,height=0.6\textheight,keepaspectratio]{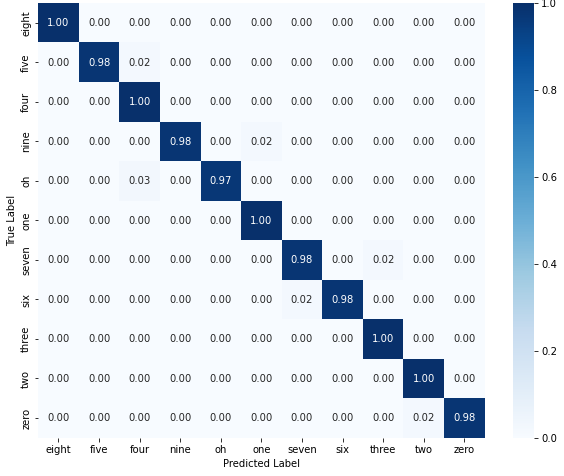} 
\caption{Confusion matrix of multiclass classification.}
\label{figure2}
\end{figure}
\begin{figure}[htbp]
\centering
\includegraphics[width=0.22\textwidth,height=0.22\textheight,keepaspectratio]{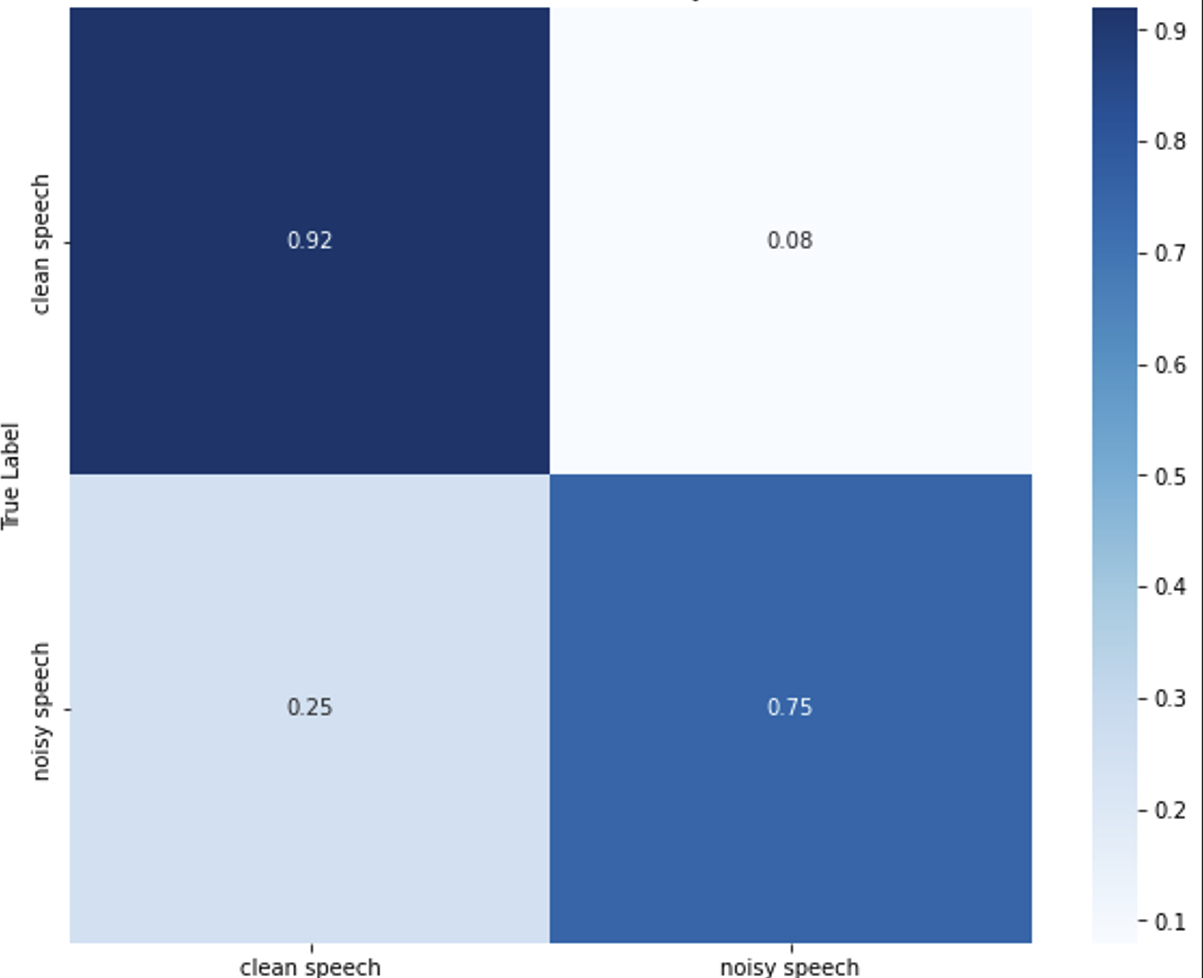} 
\caption{Confusion matrix of binary classification.}
\label{figure3}
\end{figure}

\begin{figure}[htbp]
\centering
\includegraphics[width=0.5\textwidth,height=0.6\textheight,keepaspectratio]{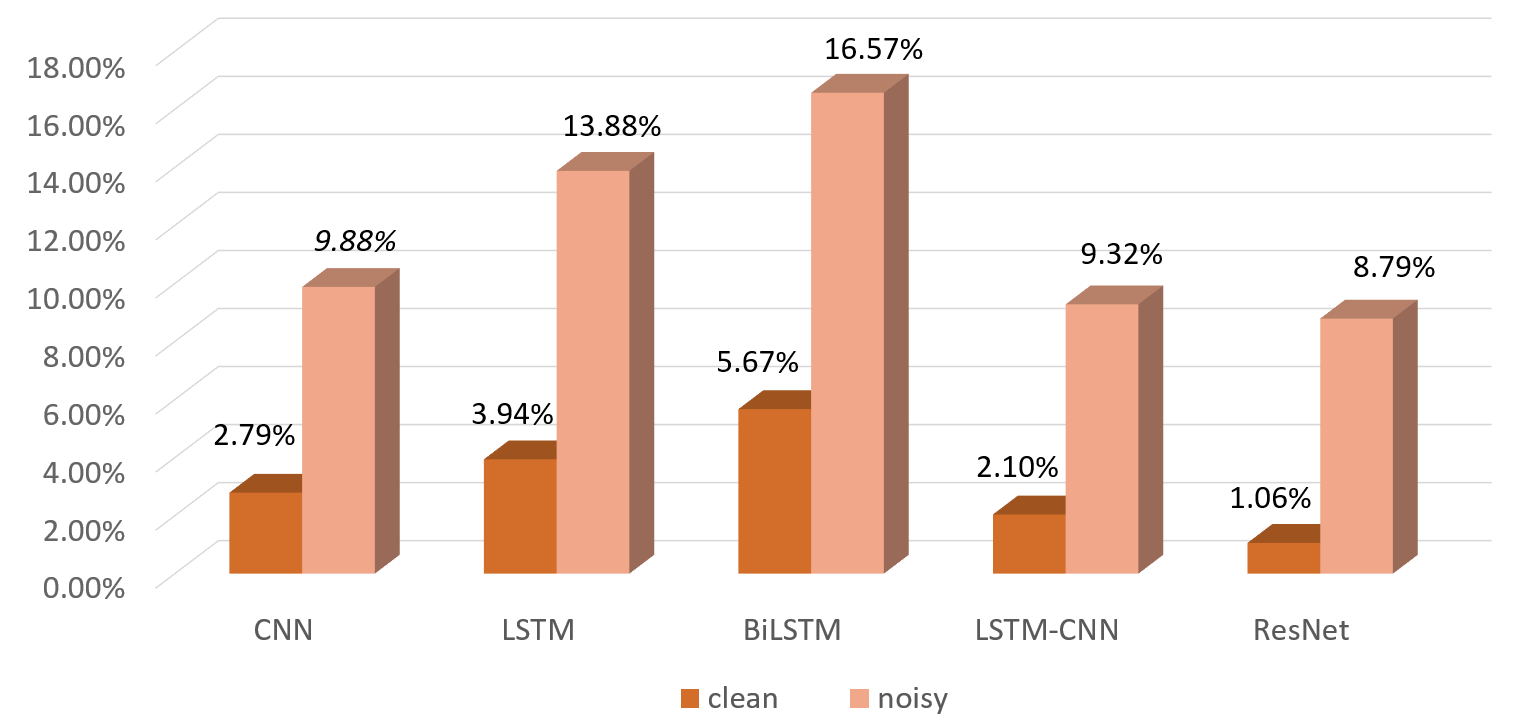} 
\caption{ WER (\%) Recognition rates obtained by CNN, LSTM, BiLSTM, and ResNet in clean and noisy mode.}
\label{figure4}
\end{figure}

\section{Conclusion}
In conclusion, our research demonstrates that ResNet-based ASR systems, when optimized with appropriate hyperparameters such as learning rate and dropout, 
achieve outstanding performance in both noisy and clean environments. Comparative analyses with CNN and LSTM models
highlight that ResNet not only matches but often exceeds their performance, establishing its robustness and reliability. The widespread application and popularity of ResNet in ASR tasks further underscore its effectiveness, confirming its superior performance in diverse and challenging environments. For futur works 
 Investigate the integration of more advanced neural network architectures, such as Transformer models \cite{kheddar2024transformers,habchi2024machine,djeffal2023automatic}, with ResNet to potentially further enhance ASR performance in noisy mode, explore advanced data augmentation techniques and noise-robust training methods to further enhance the system’s performance in extremely noisy environments.

\balance
\bibliographystyle{IEEEtran}
\begin{tiny}
\bibliography{references.bib}`

\begin{thebibliography}{10}
\providecommand{\url}[1]{#1}
\csname url@samestyle\endcsname
\providecommand{\newblock}{\relax}
\providecommand{\bibinfo}[2]{#2}
\providecommand{\BIBentrySTDinterwordspacing}{\spaceskip=0pt\relax}
\providecommand{\BIBentryALTinterwordstretchfactor}{4}
\providecommand{\BIBentryALTinterwordspacing}{\spaceskip=\fontdimen2\font plus
\BIBentryALTinterwordstretchfactor\fontdimen3\font minus \fontdimen4\font\relax}
\providecommand{\BIBforeignlanguage}[2]{{%
\expandafter\ifx\csname l@#1\endcsname\relax
\typeout{** WARNING: IEEEtran.bst: No hyphenation pattern has been}%
\typeout{** loaded for the language `#1'. Using the pattern for}%
\typeout{** the default language instead.}%
\else
\language=\csname l@#1\endcsname
\fi
#2}}
\providecommand{\BIBdecl}{\relax}
\BIBdecl

\bibitem{li2022recent}
J.~Li \emph{et~al.}, ``Recent advances in end-to-end automatic speech recognition,'' \emph{APSIPA Transactions on Signal and Information Processing}, vol.~11, no.~1, 2022.

\bibitem{essaid2024enhancing}
B.~Essaid, H.~Kheddar, and N.~Batel, ``Enhancing cochlear implant signal coding with scaled dot-product attention,'' in \emph{2024 International Conference on Telecommunications and Intelligent Systems (ICTIS)}.\hskip 1em plus 0.5em minus 0.4em\relax IEEE, 2024, pp. 1--6.

\bibitem{essaid2025deep}
B.~Essaid, H.~Kheddar, N.~Batel, and M.~E. Chowdhury, ``Deep learning-based coding strategy for improved cochlear implant speech perception in noisy environments,'' \emph{IEEE Access}, 2025.

\bibitem{kheddar2024automatic}
H.~Kheddar, M.~Hemis, and Y.~Himeur, ``Automatic speech recognition using advanced deep learning approaches: A survey,'' \emph{Information Fusion}, p. 102422, 2024.

\bibitem{zhang2017advanced}
Z.~Zhang, N.~Cummins, and B.~Schuller, ``Advanced data exploitation in speech analysis: An overview,'' \emph{IEEE Signal Processing Magazine}, vol.~34, no.~4, pp. 107--129, 2017.

\bibitem{djeffal2023automatic}
N.~Djeffal, H.~Kheddar, D.~Addou, A.~C. Mazari, and Y.~Himeur, ``Automatic speech recognition with bert and ctc transformers: A review,'' in \emph{2023 2nd International Conference on Electronics, Energy and Measurement (IC2EM)}, vol.~1.\hskip 1em plus 0.5em minus 0.4em\relax IEEE, 2023, pp. 1--8.

\bibitem{essaid2024advanced}
B.~Essaid, H.~Kheddar, N.~Batel, M.~E. Chowdhury, and A.~Lakas, ``Artificial intelligence for cochlear implants: Review of strategies, challenges, and perspectives,'' \emph{IEEE Access}, 2024.

\bibitem{noureddine2023adversarial}
K.~Noureddine, H.~Kheddar, and M.~Maazouz, ``Adversarial example detection techniques in speech recognition systems: A review,'' in \emph{2023 2nd International Conference on Electronics, Energy and Measurement (IC2EM)}, vol.~1.\hskip 1em plus 0.5em minus 0.4em\relax IEEE, 2023, pp. 1--7.

\bibitem{djeffal2023noise}
N.~Djeffal, D.~Addou, H.~Kheddar, and S.~A. Selouani, ``Noise-robust speech recognition: A comparative analysis of lstm and cnn approaches,'' in \emph{2023 2nd International Conference on Electronics, Energy and Measurement (IC2EM)}, vol.~1.\hskip 1em plus 0.5em minus 0.4em\relax IEEE, 2023, pp. 1--6.

\bibitem{he2016deep}
K.~He, X.~Zhang, S.~Ren, and J.~Sun, ``Deep residual learning for image recognition,'' in \emph{Proceedings of the IEEE conference on computer vision and pattern recognition}, 2016, pp. 770--778.

\bibitem{he2020resnet}
F.~He, T.~Liu, and D.~Tao, ``Why resnet works? residuals generalize,'' \emph{IEEE transactions on neural networks and learning systems}, vol.~31, no.~12, pp. 5349--5362, 2020.

\bibitem{barker2013pascal}
J.~Barker, E.~Vincent, N.~Ma, H.~Christensen, and P.~Green, ``The pascal chime speech separation and recognition challenge,'' \emph{Computer Speech \& Language}, vol.~27, no.~3, pp. 621--633, 2013.

\bibitem{jalalvand2015boosted}
S.~Jalalvand, D.~Falavigna, M.~Matassoni, P.~Svaizer, and M.~Omologo, ``Boosted acoustic model learning and hypotheses rescoring on the chime-3 task,'' in \emph{2015 IEEE Workshop on Automatic Speech Recognition and Understanding (ASRU)}.\hskip 1em plus 0.5em minus 0.4em\relax IEEE, 2015, pp. 409--415.

\bibitem{kinoshita2016summary}
K.~Kinoshita, M.~Delcroix, S.~Gannot, E.~A. P.~Habets, R.~Haeb-Umbach, W.~Kellermann, V.~Leutnant, R.~Maas, T.~Nakatani, B.~Raj \emph{et~al.}, ``A summary of the reverb challenge: state-of-the-art and remaining challenges in reverberant speech processing research,'' \emph{EURASIP Journal on Advances in Signal Processing}, vol. 2016, pp. 1--19, 2016.

\bibitem{nasret2021design}
A.~N. Nasret, A.~B. Noori, A.~A. Mohammed, and Z.~S. Mahmood, ``Design of automatic speech recognition in noisy environments enhancement and modification,'' \emph{Periodicals of Engineering and Natural Sciences}, vol.~10, no.~1, pp. 71--77, 2021.

\bibitem{soe2020discrete}
H.~M. Soe~Naing, R.~Hidayat, R.~Hartanto, and Y.~Miyanaga, ``Discrete wavelet denoising into mfcc for noise suppressive in automatic speech recognition system.'' \emph{International Journal of Intelligent Engineering \& Systems}, vol.~13, no.~2, 2020.

\bibitem{gueriani2024enhancing}
A.~Gueriani, H.~Kheddar, and A.~C. Mazari, ``Enhancing iot security with cnn and lstm-based intrusion detection systems,'' in \emph{2024 6th International Conference on Pattern Analysis and Intelligent Systems (PAIS)}.\hskip 1em plus 0.5em minus 0.4em\relax IEEE, 2024, pp. 1--7.

\bibitem{tsalera2021comparison}
E.~Tsalera, A.~Papadakis, and M.~Samarakou, ``Comparison of pre-trained cnns for audio classification using transfer learning,'' \emph{Journal of Sensor and Actuator Networks}, vol.~10, no.~4, p.~72, 2021.

\bibitem{rebouh2023blind}
D.~Rebouh, A.~B. Djebbar, and M.~Besseghier, ``Blind joint cfo and sto estimation for fbmc/oqam systems.'' \emph{IEEE Communications Letters}, 2023.

\bibitem{lachenani2024improving}
S.~Lachenani, H.~Kheddar, and M.~Ouldzmirli, ``Improving pretrained yamnet for enhanced speech command detection via transfer learning,'' in \emph{2024 International Conference on Telecommunications and Intelligent Systems (ICTIS)}.\hskip 1em plus 0.5em minus 0.4em\relax IEEE, 2024, pp. 1--6.

\bibitem{zhuang2020comprehensive}
F.~Zhuang, Z.~Qi, K.~Duan, D.~Xi, Y.~Zhu, H.~Zhu, H.~Xiong, and Q.~He, ``A comprehensive survey on transfer learning,'' \emph{Proceedings of the IEEE}, vol. 109, no.~1, pp. 43--76, 2020.

\bibitem{sankupellay2018bird}
M.~Sankupellay and D.~Konovalov, ``Bird call recognition using deep convolutional neural network, resnet-50,'' in \emph{Proc. Acoustics}, vol.~7, no. 2018, 2018, pp. 1--8.

\bibitem{zhang2023transfer}
L.~Zhang, Y.~Bian, P.~Jiang, and F.~Zhang, ``A transfer residual neural network based on resnet-50 for detection of steel surface defects,'' \emph{Applied Sciences}, vol.~13, no.~9, p. 5260, 2023.

\bibitem{kheddar2023deep}
H.~Kheddar, Y.~Himeur, S.~Al-Maadeed, A.~Amira, and F.~Bensaali, ``Deep transfer learning for automatic speech recognition: Towards better generalization,'' \emph{Knowledge-Based Systems}, vol. 277, p. 110851, 2023.

\bibitem{mazari2023deepTL}
A.~C. Mazari and H.~Kheddar, ``Deep learning-and transfer learning-based models for covid-19 detection using radiography images,'' in \emph{2023 International Conference on Advances in Electronics, Control and Communication Systems (ICAECCS)}.\hskip 1em plus 0.5em minus 0.4em\relax IEEE, 2023, pp. 1--4.

\bibitem{yang2023resnet}
C.~Yang, X.~Gan, A.~Peng, and X.~Yuan, ``Resnet based on multi-feature attention mechanism for sound classification in noisy environments,'' \emph{Sustainability}, vol.~15, no.~14, p. 10762, 2023.

\bibitem{mohammadamini2022learning}
M.~MohammadAmini, D.~Matrouf, J.-F. Bonastre, S.~Dowerah, R.~Serizel, and D.~Jouvet, ``Learning noise robust resnet-based speaker embedding for speaker recognition,'' in \emph{Odyssey 2022: The Speaker and Language Recognition Workshop}, 2022.

\bibitem{alzantot2019deep}
M.~Alzantot, Z.~Wang, and M.~B. Srivastava, ``Deep residual neural networks for audio spoofing detection,'' \emph{arXiv preprint arXiv:1907.00501}, 2019.

\bibitem{saleem2022e2e}
N.~Saleem, J.~Gao, M.~Irfan, E.~Verdu, and J.~P. Fuente, ``E2e-v2sresnet: Deep residual convolutional neural networks for end-to-end video driven speech synthesis,'' \emph{Image and Vision Computing}, vol. 119, p. 104389, 2022.

\bibitem{ashurov2022environmental}
A.~Ashurov, Y.~Zhou, L.~Shi, Y.~Zhao, and H.~Liu, ``Environmental sound classification based on transfer-learning techniques with multiple optimizers,'' \emph{Electronics}, vol.~11, no.~15, p. 2279, 2022.

\bibitem{tang2018acoustic}
J.~Tang, Y.~Song, L.-R. Dai, and I.~V. McLoughlin, ``Acoustic modeling with densely connected residual network for multichannel speech recognition,'' 2018.

\bibitem{reza2023customized}
S.~Reza, M.~C. Ferreira, J.~J. Machado, and J.~M.~R. Tavares, ``A customized residual neural network and bi-directional gated recurrent unit-based automatic speech recognition model,'' \emph{Expert Systems with Applications}, vol. 215, p. 119293, 2023.

\bibitem{kheddar2024transformers}
H.~Kheddar, ``Transformers and large language models for efficient intrusion detection systems: A comprehensive survey,'' \emph{arXiv preprint arXiv:2408.07583}, 2024.

\bibitem{habchi2024machine}
Y.~Habchi, H.~Kheddar, Y.~Himeur, A.~Boukabou, A.~Chouchane, A.~Ouamane, S.~Atalla, and W.~Mansoor, ``Machine learning and vision transformers for thyroid carcinoma diagnosis: A review,'' \emph{arXiv preprint arXiv:2403.13843}, 2024.

\end{thebibliography}
\end{tiny}

\end{document}